\documentclass[prl,twocolumn,superscriptaddress,amsmath,showpacs]{revtex4}
\usepackage{graphicx}

\begin{document}

\title{Directed flow in non-adiabatic stochastic pumps}
\author{Saar Rahav}
\affiliation{Department of Chemistry and Biochemistry, University of Maryland, College Park, MD 20742, USA}
\author{Jordan Horowitz}
\affiliation{Department of Physics, University of Maryland, College Park, MD 20742 USA}
\author{Christopher Jarzynski}
\affiliation{Department of Chemistry and Biochemistry, University of Maryland, College Park, MD 20742, USA}
\affiliation{Institute for Physical Science and Technology,
University of Maryland, College Park, MD 20742 USA}

\date{\today}


\begin{abstract}
We analyze the operation of a molecular machine driven by the non-adiabatic variation of external parameters.
We derive a formula for the integrated flow from one configuration to another, obtain a ``no-pumping theorem'' for cyclic processes with thermally activated transitions, and show that in the adiabatic limit the pumped current is given by a geometric expression.
\end{abstract}

\maketitle
Assemblies of molecules that perform specific tasks, known as molecular machines or motors, are ubiquitous in biological systems \cite{howard}:
they act as pumps across cell membranes, carry loads in cells and cause muscles to contract.
Remarkably, the construction and manipulation of {\it artificial} molecular machines has in recent years become feasible \cite{Reviews}.
Theoretical models, such as thermal ratchets, have provided an understanding of the microscopic mechanisms underlying the operation of molecular machines \cite{Reimann2002}.
These systems and models all exhibit the rather surprising ability to produce directed motion or current in the face of the violent thermal agitation of the mesoscopic world~\cite{Astumian2002}.

{\it Stochastic pumps} are molecular machines in which directed current is produced by varying external parameters, such as chemical environment or applied fields.
These come in two varieties: open pumps are active conduits for particle flow between reservoirs, while closed pumps generate internal currents, such as directional rotation in catenanes, mechanically interlinked ring-like molecules~\cite{Leigh2003}.
Successful theories have been developed to describe stochastic pumps under adiabatic conditions, that is when the external parameters are driven slowly~\cite{Parrondo1998,Astumian2003,Sinitsyn2007,Astumian2007},
or for weak oscillatory perturbations~\cite{Pert}.
However, a general theory of non-adiabatic pumps is lacking.
In this Letter we formulate and analyze a generic model of a non-adiabatic, closed pump, described by transitions among a network of $N$ states (e.g.\ molecular conformations).
We derive an expression for the integrated, directed current along an arbitrary link in this network (Eq.\ \ref{eq:central}) and explore its consequences.
In particular, we obtain a ``no-pumping theorem'' that applies to the cyclic variation of thermally activated transitions.

We will consider a system with a set of configurations, or {\it states}, labeled $i=1,\cdots , N$,
and will model transitions among these states as a Markov jump process.
Letting $p_i(t)$ denote the probability to be in state $i$ at time $t$, the system obeys the master equation~\cite{vanKampen}
\begin{equation}
\label{mastereq}
\dot{\bf p} = {\cal R}\, {\bf p} ,
\end{equation}
where ${\bf p} = (p_1,\cdots , p_N)$ and ${\cal R}$ is a transition matrix:
when the system is in state $j$, the probability rate to jump to state $i$ is $R_{ij}\ge 0$;
and $R_{jj} = -\sum_{i\ne j} R_{ij}$~\cite{vanKampen}.
At time $t$, the probability current from state $j$ to state $i$ is
\begin{equation}
\label{eq:Jij}
J_{ij}(t) = R_{ij}p_j - R_{ji}p_i .
\end{equation}
We imagine that the transition rates depend on a vector of external parameters $\vec\lambda$ under our direct control, i.e.\ ${\cal R} = {\cal R}({\vec\lambda})$.
For simplicity, we further assume that $R_{ij}=0$ if and only if $R_{ji}=0$; and that for any ${\vec\lambda}$, there is a unique stationary state ${\bf p}^s({\vec\lambda})$ satisfying ${\cal R} {\bf p}^s = {\bf 0}$, with stationary currents $J_{ij}^s({\vec\lambda}) = R_{ij} p_j^s - R_{ji} p_i^s$.
If $J_{ij}^s({\vec\lambda}) = 0$ for all $i, j$, then the dynamics are said to satisfy detailed balance, and we interpret the stationary distribution to be the equilibrium distribution: ${\bf p}^s({\vec\lambda}) = {\bf p}^{eq}({\vec\lambda})$.

It will prove convenient to define a {\it branching matrix} ${\cal Q}$, obtained by rescaling each column of ${\cal R}$:
$Q_{ij} = R_{ij}/\vert R_{jj}\vert$.
The diagonal elements of ${\cal Q}$ are all $-1$, and the off-diagonal elements are branching fractions:
when the system makes a transition out of state $j$, then $Q_{ij}$ is the probability that the transition is into state $i$.

\begin{figure}[th]
\center{\includegraphics[scale=0.5]{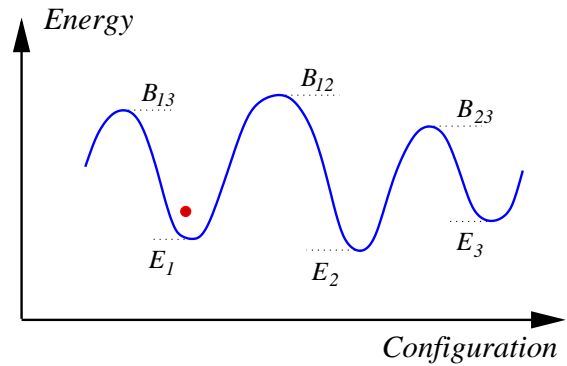}}
\caption{\label{fig1} A model stochastic pump satisfying detailed balance.
The particle makes thermal transitions among three states with energies $E_j(\vec\lambda)$, over barriers with energies $B_{ij}(\vec\lambda)$.
These are varied with time to induce currents.}
\end{figure}

As an illustrative example, for which the dynamics satisfy detailed balance, imagine thermally activated transitions between potential wells separated by energetic barriers, with well depths $E_j(\vec\lambda)$ and barrier energies $B_{ij}(\vec\lambda)$ (Fig.~\ref{fig1}).
We then have $p_j^s = p_j^{eq} \propto e^{-\beta E_j}$, and transition rates take the familiar Arrhenius form \cite{Astumian2007,Arrhenius_comment}
$R_{ij} = k \exp\left[ -\beta (B_{ij} - E_j) \right]$,
from which it follows that the branching fractions depend on the barrier energies but not on the well depths:
\begin{equation}
\label{eq:qb}
Q_{ij} = e^{-\beta B_{ij}} / \sum_{k\ne j} e^{-\beta B_{kj}}.
\end{equation}
We will use this result in our later analysis.

We will consider a process during which the system evolves as the parameters are varied externally, ${\vec\lambda}={\vec\lambda}_t$, from ${\vec\lambda}_0=A$ to ${\vec\lambda}_\tau=B$.
For a fixed but arbitrary pair of states $m$, $n$, the {\it integrated current},
$\Phi_{mn} \equiv \int_0^\tau\mathrm{d}t \, J_{mn}(t)$, represents the net transfer of probability from $n$ to $m$.
This is a measure of the directed motion produced along this particular link in the network of states.
We will show that the integrated current is given by the compact expression,
\begin{equation}
\label{eq:central}
\Phi_{mn} = \int_0^\tau dt \,
\left[ J_{mn}^s({\vec\lambda}_t) + {\bf V}_{mn}({\vec\lambda}_t) \cdot \dot{\bf p}(t) \right] ,
\end{equation}
where the vector field ${\bf V}_{mn}({\vec\lambda}) = (V_{mn,1}, \cdots , V_{mn,N})$ is given by Eq.~\ref{eq:V_explicit} below.
Since in general we must integrate Eq.~\ref{mastereq} to obtain $\dot{\bf p}(t)$, Eq.~\ref{eq:central} is not a shortcut for calculating $\Phi_{mn}$.
Rather, it provides a useful and non-trivial decomposition of $\Phi$ into a {\it stationary} contribution, $\Phi^{s} = \int J^s \, dt$, due to a baseline current that flows even at fixed ${\vec\lambda}$; and the remaining {\it excess}, or pumped contribution, $\Phi^{ex}$, due to the redistribution of probabilities induced by the variation of ${\vec\lambda}$.
The latter is the contraction of a path ${\bf p}(t)$ along the field ${\bf V}_{mn}(\vec\lambda)$:
\begin{equation}
\label{eq:contraction}
\Phi_{mn}^{ex} = \int {\bf V}_{mn}(\vec\lambda_t)  \cdot d{\bf p}(t).
\end{equation}
While Eq.~\ref{eq:central} is valid quite generally, we note two consequences that follow in particular situations.

(1) When ${\vec\lambda}$ is varied slowly, the system remains near the stationary state, ${\bf p}(t)\approx{\bf p}^s(\vec\lambda_t)$.
This suggests we replace $d{\bf p}$ in Eq.~\ref{eq:contraction} by $d{\vec\lambda} \circ \vec\nabla{\bf p}^s \equiv \sum_\mu d\lambda_\mu (\partial {\bf p}^s/\partial\lambda_\mu)$~\cite{explain_circ},
in the adiabatic limit.
With this replacement, which can be justified by appeal to an adiabatic perturbation theory \cite{adb}, we get
\begin{equation}
\label{eq:Phi_ex_ad}
\Phi_{mn}^{ex} = \int_A^B \vec A_{mn}(\vec\lambda) \circ  d{\vec\lambda}  ,
\end{equation}
where $\vec A_{mn}(\vec\lambda) \equiv {\bf V}_{mn} \cdot \vec\nabla {\bf p}^s$.
This expression is {\it geometric}: time no longer appears here, and $\Phi_{mn}^{ex}$ is simply a line integral of $\vec A_{mn}$ along a path in parameter-space.
Similar geometric results have been obtained by Astumian for a three-state system~\cite{Astumian2007}, and by Sinitsyn and Nememan~\cite{Sinitsyn2007} for open stochastic pumps.

(2) When ${\cal R}$ describes thermally activated (Arrhenius) transitions over barriers, with parameter-dependent well depths and barrier energies (Fig.~\ref{fig1}), then we will show that Eq.~\ref{eq:central} leads to a surprising ``no-pumping theorem" for cyclic processes: $\Phi^{ex}=0$ if the well depths are varied while the barrier energies are held fixed, or (trivially) vice-versa.
Only by varying {\it both well depths and barrier energies} during a pumping cycle, can we generate a net transfer of probability between states.

Eq.~\ref{eq:central} is derived by eliminating ${\bf p}$ from Eqs.~\ref{mastereq} and \ref{eq:Jij} to obtain $J_{mn}(t) = J_{mn}^s + {\bf V}_{mn}\cdot \dot{\bf p}$.
We now sketch the steps of this derivation, skipping tedious but routine linear-algebraic manipulations.
The final results give ${\bf V}_{mn}$ in terms of readily evaluated minors of the matrices ${\cal R}$ (Eq.~\ref{eq:V_explicit}) or ${\cal Q}$ (Eq.~\ref{eq:vij_db_q}).
We note before proceeding that there is a gauge freedom at play here: since $\sum_j \dot p_j = 0$ by conservation of probability, Eq.~\ref{eq:central} is unaffected by the replacement ${\bf V}_{mn} \rightarrow {\bf V}_{mn} + f(\vec\lambda) {\bf 1}$, where ${\bf 1} \equiv (1,1,\cdots , 1)$ and $f$ is an arbitrary function.
Thus our results for ${\bf V}_{mn}$ are not unique, merely convenient.

Eq.~\ref{mastereq} is a set of linear equations, which we label $\hat e_1, \cdots, \hat e_N$.
Since $\det {\cal R} = 0$, these are linearly dependent (one of them is redundant) and ${\cal R}$ cannot be inverted to solve for ${\bf p}$ in terms of $\dot{\bf p}$.
Specifically, for a given $\dot{\bf p}$,
if ${\bf p}$ satisfies Eq.~\ref{mastereq} then so does
${\bf p}^{(\alpha)} = {\bf p} + \alpha {\bf p}^s$,
for any value of $\alpha$.
We remove this degeneracy
by imposing the normalization condition ${\bf 1}\cdot{\bf p} = 1$,
which we label $\hat e^\prime$: replacing $\hat e_N$ by $\hat e^\prime$ in Eq.~\ref{mastereq}, we get a set of linearly {\it independent} equations
\begin{equation}
\label{eq:independent}
\dot{\bf p}^\prime = {\cal R}^\prime \, {\bf p} ,
\end{equation}
where
$\dot{\bf p}^\prime \equiv (\dot p_1, \cdots, \dot p_{N-1}, 1)$,
and ${\cal R}^\prime$ is obtained by substituting the vector ${\bf 1}$ for the $N$'th row of ${\cal R}$.
Since $\det{\cal R}^\prime \ne 0$, we solve for ${\bf p}$ using Cramer's rule~\cite{LinAl}:
\begin{equation}
\label{Pgeneral}
 p_j = \det {\cal R}_j^\prime / \det {\cal R}^\prime,
 \end{equation}
 where ${\cal R}_j^\prime$ is obtained from ${\cal R}^\prime$ by replacing the $j$'th column by $\dot{\bf p}^\prime$.
Expanding $\det {\cal R}_j^\prime$ along this column (for $j=m,n$), then substituting Eq.~\ref{Pgeneral} into Eq.~\ref{eq:Jij}, we get
 \begin{equation}
 \label{eq:Jlin}
 J_{mn} = \sum_{k=1}^N (-1)^k
 \left( \sigma_{mn,k} - \sigma_{nm,k} \right) \dot p_k^\prime.
 \end{equation}
 Here $\sigma_{ij,k}(\vec\lambda) = (-1)^j R_{ij} \, r_j^\prime(k;j) / r^\prime$, where $r^\prime = \det{\cal R}^\prime$,
and $r_i^{\prime}(a;b)$ denotes the $(a;b)$ minor of ${\cal R}_i^\prime$, that is the determinant of the matrix obtained by deleting row $a$ and column $b$ of ${\cal R}_i^\prime$.
Comparing Eq.~\ref{eq:Jlin} with the integrand in Eq.~\ref{eq:central}, and recognizing that $J_{mn} = J_{mn}^s$ when ${\bf p} = {\bf p}^s$ (i.e.\ when $\dot{\bf p} = 0$), we obtain
\begin{equation}
\label{eq:V_explicit}
V_{mn,k} = (-1)^k \, (1 - \delta_{kN}) \, (\sigma_{mn,k} - \sigma_{nm,k})  ,
\end{equation}
which gives ${\bf V}_{mn}(\vec\lambda)$ in terms of the elements of ${\cal R}(\vec\lambda)$.

Let us now separately analyze the case in which the dynamics satisfy detailed balance, for all $\vec\lambda$.
Recalling that Eq.~\ref{mastereq} supports a family of solutions, ${\cal F} = \{ {\bf p}^{(\alpha)} = {\bf p} + \alpha {\bf p}^s \}$,
we note that if we formally replace ${\bf p}$ by ${\bf p}^{(\alpha)}$ in Eq.~\ref{eq:Jij}, we obtain $J_{ij}^{(\alpha)} = J_{ij} + \alpha J_{ij}^s$.
In general, $J_{ij}^{(\alpha)}$ is not a physically meaningful quantity.
However, if ${\cal R}$ satisfies detailed balance, as we assume in this paragraph, then $J_{ij}^s = 0$,
and therefore {\it all solutions ${\bf p}^{(\alpha)} \in {\cal F}$ give the same current}, $J_{ij}^{(\alpha)} = J_{ij}$, upon substitution into Eq.~\ref{eq:Jij}.
We now exploit this observation: rather than solving for the vector ${\bf p} \in {\cal F}$ that satisfies normalization ($\hat e^\prime$), as done above, we instead choose the vector $\bar{\bf p} \in {\cal F}$ that satisfies $\bar p_m = 0$ (we label this condition $\hat e^{\prime\prime}$) and then we substitute $\bar{\bf p}$ into Eq.~\ref{eq:Jij} to determine $J_{mn}$.
This approach leads to an expression for ${\bf V}_{mn}$ that (unlike Eq.~\ref{eq:V_explicit}) depends only on the branching fractions ${\cal Q}$ and not on the transition rates ${\cal R}$.
In detail, to solve for $\bar{\bf p}$ we break the degeneracy of Eq.~\ref{mastereq} by replacing $\hat e_m$ with $\hat e^{\prime\prime}$, rather than $\hat e_N$ with $\hat e^\prime$ as earlier.
In lieu of Eq.~\ref{eq:independent} we now have
$\dot {\bf p}^{\prime\prime} = {\cal R}^{\prime\prime} \, \bar{\bf p}$,
where
$\dot{\bf p}^{\prime\prime} = (\dot p_1, \cdots, \dot p_{m-1}, 0, \dot p_{m+1}, \cdots, \dot p_N)$,
and ${\cal R}^{\prime\prime}$ is defined by replacing the $m$'th row of ${\cal R}$ with $(0\cdots 0\,1\,0 \cdots 0)$, where the 1 is on the diagonal.
Now,
(1) using Cramer's rule to solve for $\bar{\bf p}$, then
(2) taking $J_{mn} = R_{mn} \bar p_n$ (since $\bar p_m=0$), and
(3) recalling that $Q_{ij} = R_{ij}/\vert R_{jj}\vert$,
after some effort we obtain
\begin{subequations}
\label{eq:pdb}
\begin{eqnarray}
\label{pumpeddb}
\Phi_{mn} &=&
\int_0^\tau dt\, {\bf V}_{mn}(\vec\lambda_t) \cdot \dot{\bf p}(t) \\
\label{eq:vij_db_q}
V_{mn,k} &=& (-1)^{n+k} \, \Lambda \, (1-\delta_{mk}) \, \frac{Q_{mn} \, q(m,k;m,n)}{q(m;m)} .
\end{eqnarray}
\end{subequations}
Here $q(m;m)$ is the $(m;m)$ minor of ${\cal Q}$, and similarly
$q(m,k;m,n)$ is the determinant obtained after deleting rows $m$ and $k$ and columns $m$ and $n$ of ${\cal Q}$;
finally, $\Lambda = -1$ if $n < m < k$ or $k < m < n$, otherwise $\Lambda=+1$.

Now let us use Eq.~\ref{eq:pdb} to establish a ``no-pumping theorem'' pertaining to cyclic processes.
We continue to assume that ${\cal R}(\vec\lambda)$ satisfies detailed balance for all $\vec\lambda$, and we picture the dynamics as arising from thermally activated transitions over barriers, with externally controlled well depths and barrier energies (Fig.~\ref{fig1}).
We now imagine a process during which the parameters are held fixed at $\vec\lambda_A$ prior to $t=0$; then during the interval $0\le t \le T$ they are made to trace out a closed loop in parameter space, after which they are again held fixed at $\vec\lambda_A$.
In this scenario, from the distant past to the distant future the vector ${\bf p}(t)$ also evolves through a closed path, returning to its initial equilibrium state: ${\bf p}(\pm\infty) = {\bf p}^{eq}(\vec\lambda_A)$.
We are interested in the integrated current, $\Phi_{mn} = \int_{-\infty}^{+\infty} dt\, J_{mn}$ during such a process.
Let us separately consider two cases.
First, if only the barrier energies (and not the well depths) are varied, then the system simply remains in equilibrium at all times, ${\bf p}(t) = {\bf p}^{eq}(\vec\lambda_A)$, and there are no currents.
Now consider the less obvious case in which the barrier energies are held fixed, but non-zero currents $J_{mn}(t)$ are generated through a cyclic variation of the well depths.
Since ${\bf V}_{mn}$ is determined by the elements of ${\cal Q}$ (Eq.~\ref{eq:vij_db_q}), which in turn depend only on the fixed barrier energies (Eq.~\ref{eq:qb}), Eq.~\ref{pumpeddb} becomes $\Phi_{mn} = {\bf V}_{mn} \cdot \int_{-\infty}^{+\infty} dt \, \dot {\bf p}(t) = 0$.
Thus, in order to generate non-zero integrated current during a cyclic process, {\it both well depths and barrier energies must be varied}.

This result extends to encompass repeated, periodic cycling of the parameters,
$\vec\lambda(t+\tau) = \vec\lambda(t)$,
in which case Floquet theory ensures that the system relaxes to a time-periodic state,
${\bf p}(t+\tau) = {\bf p}(t)$~\cite{Talkner1999}.
The arguments of the previous paragraph then apply to this periodic state: if the barrier energies are held fixed, then ${\bf V}_{mn}(\vec\lambda_t)$ is constant in time, and therefore Eq.~\ref{pumpeddb}
gives $\Phi_{mn}=0$ over one period of driving.

We have just argued that all $\Phi_{mn}$'s vanish in a cyclic process during which ${\cal Q}$ is fixed and detailed balance is satisfied.
It is of interest to count the number of constraints represented by these conditions, and to compare this with the general number of constraints needed to ensure absence of directed flow.
Assuming $E$ links among our network of $N$ states, there are $2E$ non-zero branching fractions $Q_{ij}$.
Such a network can be decomposed into $E-N+1$ closed loops, with each link belonging to at least one loop~\cite{Schnakenberg1976}.
The $2E$ non-zero branching fractions are not independent: conservation of probability imposes $N$ conditions $\sum_i Q_{ij}=0$ ($j=1, \cdots , N$), and detailed balance imposes an additional $E-N+1$ conditions, representing the absence of thermodynamic force around each closed loop~\cite{Andrieux2007}.
Therefore by choosing $E-1$ specific branching fractions to be $\vec{\lambda}$-independent, we guarantee that the others will also be $\vec{\lambda}$-independent.
On the other hand, we can ensure that all $\Phi_{mn}$'s vanish by insisting that there be zero integrated current around every closed loop in the network decomposition.
Since there are $E-N+1$ such loops, it is clear that fixing the branching fractions (i.e.\ imposing $E-1$ constraints) is not the most general condition for the absence of integrated flow.
However, it is a simple condition, which has the same leading order scaling for systems with many states, assuming $E \propto N^2$.

We now illustrate our results using a model system, motivated by an experiment by Leigh  {\em et.\ al.\ }\cite{Leigh2003} and analyzed in the adiabatic limit by Astumian \cite{Astumian2007}.
We consider thermally activated transitions among three states, depicted by the wells in Fig.~\ref{fig1}, with rates $R_{ij} = k \exp\left[ -\beta (B_{ij} - E_j) \right]$, and we will take $k, \beta=1$ to set the units of time and energy.
${\cal R}$ satisfies detailed balance, but by varying the well depths and barrier energies we can induce non-zero currents.
Recalling Eq.~\ref{eq:qb}, and defining $\psi_1 = \exp(-B_{12}-B_{13})$, $\psi_2 = \exp(-B_{12}-B_{23})$, $\psi_3 = \exp(-B_{13}-B_{23})$, and $K = \sum_j \psi_j$, we evaluate Eq.~\ref{eq:vij_db_q} for $(m,n)=(2,1)$ to obtain
\begin{equation}
{\bf V}_{21} = K^{-1} (-\psi_1-\psi_2, 0 , -\psi_1) \rightarrow K^{-1} (-\psi_2 , \psi_1 , 0),
\end{equation}
where in the last step we have used the gauge freedom ${\bf V}_{21} \rightarrow {\bf V}_{21} + (\psi_1/K) \, {\bf 1}$.
 
When $\vec\lambda$ is varied adiabatically around a closed path, the pumped current is given by Eq.~\ref{eq:Phi_ex_ad}, with $\vec A_{21} = (-\psi_2 \vec\nabla p_1^{eq} + \psi_1 \vec\nabla p_2^{eq})/K$.
If the barrier energies are held fixed during this process, then $\psi_1$, $\psi_2$, and $K$ are constant, hence the integrand is a total differential and  there is no pumped current, as predicted in Ref.~\cite{Astumian2007}.

\begin{figure}[th]
\center{\includegraphics[scale=0.32]{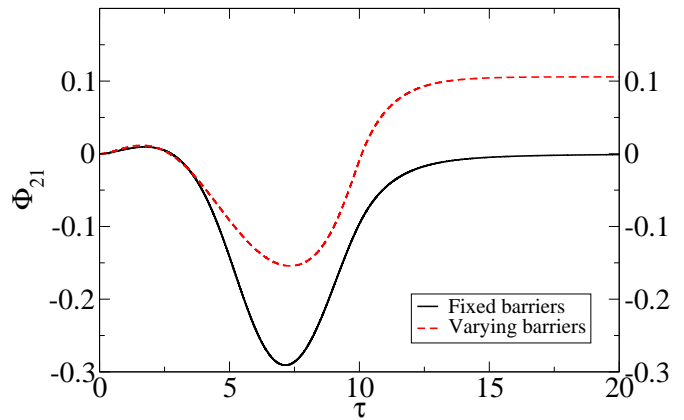}}
\caption{\label{fig2} The integrated current $\Phi_{21}$ for
non-adiabatic cycles with fixed barriers (solid line) or varying
barriers (dashed).}
\end{figure}

Now let us analyze cyclic but {\it non-adiabatic} variation of the well depths and barrier energies.
We first consider a process during which the barriers are held fixed.
Specifically, we take $(B_{12}, B_{23},B_{13}) = (-0.3,0.5,0)$, and
\begin{equation}
\label{eq:protocol}
E_j(t) = -2 +  \cos \left[ 2 \pi \left( \frac{t}{T} + \frac{j-1}{3} \right) \right] ,
\end{equation}
for $0 < t < T=10$.
Thus the well depths $E_j(t)$ undergo one cycle of pumping, with phases staggered by $2\pi/3$ in a piston-like sequence.
Outside this time interval all parameters are fixed, so the system ultimately relaxes to its initial equilibrium state.
The solid line in Fig.~\ref{fig2} shows the integrated current $\Phi_{21}(\tau) = \int_0^\tau dt \, J_{21}(t)$, obtained by numerical integration of Eqs.~\ref{mastereq} and \ref{eq:Jij}.
We see that probability sloshes back and forth on the link between states $1$ and $2$: initially there is a gentle flow from $1$ to $2$ ($d\Phi_{21}/d\tau>0$ for  $\tau\lesssim 2$), then an interval of stronger current in the opposite direction, followed by another reversal shortly before $\tau=7.5$.
The eventual decay of $\Phi_{21}$ to zero indicates a net cancellation of these flows, as predicted by our no-pumping result.
We next consider a process during which both well depths and barrier energies are varied with time: the $E_j$'s are again driven according to Eq.~\ref{eq:protocol}, but now each barrier moves in synchrony with the well to its immediate right in Fig.~\ref{fig1};
e.g.\ as $E_1$ goes down and then up, so does $B_{13}$, so that their difference remains fixed at $B_{13}-E_1 = 2 = B_{12}-E_2=B_{23}-E_3$.
The integrated current $\Phi_{21}(\tau)$ is shown by the dashed line in Fig.~\ref{fig2}; the asymptotic value $\Phi_{21} \approx 0.1$ reveals a net transfer of probability from state 1 to state 2 over the cycle.
Note that in both cases non-vanishing currents persist for some time after $\tau=T$, reflecting the decay to equilibrium that occurs after the parameters stop being varied.

Our no-pumping theorem applies to a single particle jumping among potential wells (Fig.~\ref{fig1}).
When more particles are present, and they interact with one another,
the general results Eqs.~\ref{eq:central}, \ref{eq:V_explicit}, and \ref{eq:vij_db_q} remain valid, but now the roles of wells and barriers are played by many-body energies.
We then find non-zero cyclic currents even when the single particle barriers are held fixed~\cite{adb}.
These results are consistent with the experimental observation of currents in 3- but not 2-catenanes~\cite{Leigh2003};
these linked-ring molecules are naturally modeled as two- or one-particle systems, respectively, with the particle(s) jumping among binding sites whose affinities are varied externally~\cite{Astumian2007}.
This agreement suggests that the non-adiabatic framework described in this Letter will prove useful in the design and analysis of nanoscale stochastic pumps.

Recently, we have learned of a generalized no-pumping theorem derived by Chernyak and Sinitsyn \cite{Chernyak2008}.  
Their results apply to systems satisfying detailed balance and account for the topology of the network.

We gratefully acknowledge Doran Bennett and Nikolai Sinitsyn for useful discussion and correspondence, and the University of Maryland for financial support.


\begin{thebibliography}{99}

\bibitem{howard}
J. Howard, {\em Mechanics of Motor Proteins and the Cytoskeleton}
(Sinauer, Sunderland, 2001).

\bibitem{Reviews}
G. S. Kottas {\it et al}., Chem. Rev. {\bf 105}, 1281 (2005);
W. R. Browne and B. L. Feringa, Nat. Nanotechnol. {\bf 1}, 25 (2006);
E. R. Kay, D. A. Leigh, and F. Zerbetto, Angew. Chem. Int. Ed. {\bf 46}, 72 (2007);
B. L. Feringa, J. Org. Chem. {\bf 72}, 6635 (2007).

\bibitem{Reimann2002}
P. Reimann,  Phys. Rep. {\bf 361}, 57 (2002).

\bibitem{Astumian2002}
R. D. Astumian and P. H\"anggi, Phys. Today {\bf 55}, No. 11, 33-39 (2002).

\bibitem{Leigh2003}
D. A. Leigh {\it et al}., Nature
{\bf 424}, 174 (2003).

\bibitem{Parrondo1998}
J. M. R. Parrondo, \pre {\bf 57}, 7297 (1998).

\bibitem{Astumian2003}
R. D. Astumian, \prl {\bf 91}, 118102 (2003).

\bibitem{Sinitsyn2007}
N. A. Sinitsyn and I. Nemenman, Europhys. Lett. {\bf 77},
58001 (2007); \prl {\bf 99}, 220408 (2007).

\bibitem{Astumian2007}
R. D. Astumian, Proc. Nat. Acad. Sci. {\bf 104}, 19715 (2007).

\bibitem{Pert}
I. Sokolov, J. Phys. A: Math. Gen. {\bf 32}, 2541 (1999);
R. D. Astumian and I. Derenyi, \prl {\bf 86}, 3859 (2001);
K. Jain {\it et al}., \prl {\bf 99}, 190601 (2007);
J. Ohkubo, J. Stat. Mech., P02011 (2008).

\bibitem{vanKampen}
N. G. van Kampen, {\em Stochastic Processes in Physics and Chemistry} (New York, Elsevier, 2007).

\bibitem{Arrhenius_comment}
We implicitly assume that intra-well relaxation occurs on a time scale faster than that of the external driving. See B. Caroli {\em et al}., Physica A {\bf 108}, 233 (1981).

\bibitem{explain_circ}
For clarity, different symbols $\cdot$ and $\circ$ are used to denote dot products in ${\bf p}$-space and $\vec\lambda$-space, respectively.

\bibitem{adb}
J. Horowitz, S. Rahav, and C. Jarzynski (unpublished).

\bibitem{LinAl}
S. J. Leon, {\em Linear Algebra with Applications} (New Jersey, Prentice Hall, 2002).

\bibitem{Talkner1999}
P. Talkner, New J. Phys. {\bf 1}, 4.1 (1999).

\bibitem{Schnakenberg1976}
J. Schnakenberg, \rmp {\bf 48}, 571 (1976).
We use the term ``loop'' in place of Schnakenberg's 	``cycle'', to avoid confusion with our previous use of the latter.

\bibitem{Andrieux2007}
D. Andrieux and P. Gaspard, J. Stat. Phys. {\bf 127}, 107 (2007).

\bibitem{Chernyak2008}
V. Y. Chernyak and N. A. Sinitsyn, private communication.
\end{thebibliography}
\end{document}